\documentclass[twocolumn]{aa}

\renewcommand{\b}{\begin{eqnarray}}
\renewcommand{\d}{\end{eqnarray}}

\usepackage{natbib}
\bibpunct{(}{)}{;}{a}{}{,}
\usepackage{graphicx}
\usepackage[mathcal]{eucal}
\newcommand{\ds}{\displaystyle}
\newcommand{\inv} {\frac {1}}

\newcommand{\deriv} [2] {\frac {d #1 } {d #2} }

\newcommand{\eqna} [1] {
\begin{eqnarray}#1
\end{eqnarray}}

\usepackage{txfonts}
\begin{document}
\title{A closure model with plumes}

   \subtitle{II. Application to the stochastic excitation of solar $p$~modes}
    \author{K.~Belkacem\inst{1} 
            \and
            R.~Samadi\inst{1} 
            \and
            M.~J. Goupil\inst{1} 
            \and
            F.~Kupka\inst{2}
            \and
            F.~Baudin\inst{3}}

 \institute{
Observatoire de Paris, LESIA, CNRS UMR 8109, 92195 Meudon, France \and
Max-Planck-Institute for Astrophysics, Karl-Schwarzschild Str. 1, 85741 Garching, Germany \and
Institut d'Astrophysique Spatiale, CNRS/Universit\'e Paris XI UMR 8617,91405 Orsay Cedex, France }
\offprints{K. Belkacem}
\mail{Kevin.Belkacem@obspm.fr}
\date{
Received 06 April 2006 / Accepted 30 June 2006}

\titlerunning{A closure model with plumes}

  \abstract{ Amplitudes of stellar $p$~modes result
  from a balance between excitation and damping processes taking place in
  the uppermost part of convective zones in solar-type stars and
  can therefore be used as a seismic diagnostic for the physical properties of these external layers.
   }
   { Our  goal is to improve the theoretical modelling of stochastic excitation of $p$~modes by turbulent convection.
   }
   { With the help of the closure model with plume (CMP) developed in
   a companion paper, we refine the theoretical description of the excitation by the turbulent Reynolds stress term.
   The CMP is generalized for two-point correlation products so as to apply it to
   the formalism developed by Samadi \& Goupil (2001). The excitation source terms are
   then computed with this improvement, and a comparison with solar data from the GOLF instrument is performed.
   }
   {The present model provides a significant improvement when comparing absolute values of theoretical ampplitudes with observational data. It  
   gives rise to  a frequency dependence of the power supplied to solar $p$~modes, which agrees with GOLF
   observations. It is shown that the asymmetry of the turbulent convection zone (up and downflows)
   plays a major role in the excitation processes. Despite an increase in the Reynolds stress term contribution 
   due to our improved description, an additional source of excitation, identified as the entropy source term, is still necessary for reproducing the 
   observational data. 
   }
   {Theoretical excitation rates in the
   frequency range $\nu \in [2.5 \, {\rm mHz}, 4 \, {\rm mHz}]$ now are in agreement with the observational data from the GOLF instrument.
   However, at lower frequencies, it exhibits small discrepancies at the maximum level of a few per cent. Improvements are likely to come from 
   a better physical description of the excitation by entropy fluctuations in the superadiabatic zone.
   }

   \maketitle
%
%________________________________________________________________

\section{Introduction}
Amplitudes of solar-like oscillations result from a balance between excitation and damping. Excitation is attributed to turbulent 
motions that excite the $p$~modes. 
In the uppermost part of the convection zone, entropy fluctuations and eddy motions drive oscillations. 
In this region, convection becomes inefficient and there is an increase 
of in the eddy velocities and entropy fluctuations. Solar-like oscillations are mainly excited in such a region, 
thus a theoretical model of the excitation processes is a powerful 
tool in understanding the properties of the convective zones of solar-type stars.  
\cite{GK77} have proposed a model for the excitation process using the turbulent Reynolds stress and deducing 
an estimation of the power supplied to the $p$~modes. 
The underestimation of the excitation rates by around a factor $10^3$ compared to the observed solar values \citep{Osaki90} 
led to alternative formulations \citep{Gold90,GK94}. 
Another source of excitation was identified  by \cite{GK94}:  the so-called entropy source term.
Its contribution cannot be neglected, even though \cite{Stein01B} have shown that excitation from the Reynolds stress remains dominant 
in comparison with the entropy fluctuation source term.\\
\cite{Samadi00I} propose a generalized formalism, taking the Reynolds and entropic fluctuation source terms into account. 
This approach allows investigation of the effects of several models of turbulence \citep{Samadi02I,Samadi02II} by expressing the source 
terms as functions of the turbulent kinetic energy spectrum and the time-correlation function. \\
A confrontation of this model with data from the BiSON instrument \citep[data from][]{Chaplin98} led to the conclusion that the theoretical predictions 
were in good agreement with the observations \citep{Samadi02II}. 
Nevertheless, observational data from the GOLF instrument and a study of the BiSON data  indicate  that some discrepancies remain  
between the theoretical computation  and  observational  data.
In \cite{Samadi00I} (see also Samadi et al., 2005),  one of the main assumptions is the quasi-normal approximation (QNA), 
which is useful for correlation functions of the turbulent Reynolds stress and the entropy fluctuation source terms \citep{Samadi00I}. \\
The uppermost part of the convection zone being a turbulent convective system composed of two flows, the probability distribution 
function of the fluctuations of the vertical
 velocity and temperature does not follow a Gaussian law \citep{Lesieur97}. Thus, the use of the QNA, which is exact 
for a normal distribution, becomes a doubtful approximation.\\
In a companion paper \citep[][hereafter Paper I]{Belkacem06a}, we propose another approach in order  to build a closure model 
that expresses  fourth-order correlation functions in terms of the second-order ones. 
This alternative approach  consists in considering the convection zone as composed of two flows 
(the updrafts and downdrafts). Starting from the \citet{GH2002} approach, we develop a generalized two-scale mass-flux model (GTFM) that takes  
the physical properties of each flow into account. 
Then a theoretical description of the plumes developed by \citet{RZ95} is used to  construct the closure model 
with plumes (CMP). This model is valid \
for one-point correlation functions and in the quasi-adiabatic zone. 
However, what is needed here is a closure model for two point correlation functions.  In the present  paper,  we then propose a simple way 
to obtain this closure model to use it for calculating of the excitation rates according to  \cite{Samadi00I}.
Only the Reynolds stress source term  is corrected, mainly because it is  the dominant term    \citep{Stein01B,Samadi02II}.
The entropy fluctuations 
are considered in the same way as explained in \cite{Samadi00I} (i.e. using the QNA approximation).\\
The paper is organized as follows: the theoretical model of stochastic excitation of $p$~modes is briefly summarized in Sect. 2.
In Sect. 3, the closure model with plume (CMP) is generalized for two-point correlation products and implemented into the formalism 
of \cite{Samadi00I}. In 
 Sect. 4, the calculation of theoretical power is explained. In Sect. 5,  GOLF observational data 
are presented together with the derivation of observable quantities. 
A comparison between the theoretical power and heights computed  as described in Sect. 4 with the
corresponding observed quantities defined in Sect. 5 is performed in Sect. 6.  
Section 7 is dedicated to discussions and conclusions.\\

%__________________________________________________________________

\section{A model for stochastic excitation of solar-like  $p$ modes}
\label{MSE}

The theoretical model of stochastic excitation considered here
 is basically that of \citet{Samadi00I} (see also Samadi et al.,  2005).
It takes two sources into account that drive the resonant modes of the stellar cavity. The
first one is related to the Reynolds stress tensor and as such
represents a mechanical source of excitation.  The second one is
caused by the advection of the turbulent fluctuations of entropy by
the turbulent motions (the so-called ``entropy source term'') and thus
 represents a thermal source of excitation \citep{GK94,Stein01B}.

The power fed into each mode, $P$, is given by \citep[see e.g.][]{Samadi00II}:
\begin{equation}
\label{power}
P\equiv \deriv{E}{t}   =  2 \eta\, E =  \eta\,  {\langle |A|^2 \rangle}\,I\,\omega_0^2 \;,
\end{equation}
where $\langle \rangle$ denotes the ensemble average, $\langle  |A| ^2 \rangle $ the mean square amplitude, $\eta$ the damping rate, and $E$ the energy that is defined as
\begin{equation}
E=  \inv {2} ~\langle |  A | ^2 \rangle  \; I \omega_0^2
\end{equation}
where $\ds{I}$ is the mode inertia and $\omega_0$ is the oscillation eigenfrequency \citep[see][for details]{Samadi00I}.\\
The mean square amplitude, as explained in \citet{Samadi00I}, is
\begin{equation}
\label{A2}
\left < \mid A \mid ^2 \right >  = \frac{1}{8 ~\eta ~(I \omega_0)^2}
\left ( C_R^2 + C_S^2  \right )
\end{equation}
where $C_R^2$ and $C_S^2$ are the turbulent Reynolds stress and entropy contributions, respectively.
Their expressions for radial modes are given by
\eqna{
C_R^2  &= &  \int  d^3 x_0 \, \rho_0^2 ~f_r ~  \int_{-\infty}^{+\infty} d\tau \, e^{-i\omega_0 \tau} \, \int d^3r \, \left <
 w_1^2  w_2^2
\right >   
\label{C2R_rad} \\
C_S^2  &= &   \int  d^3x_0  \, g_r  \int_{-\infty}^{+\infty} d\tau \, e^{-i\omega_0 \tau}
\int d^3r \, \left <
 \left ( w s_t     \right)_1
\left ( w s_t  \right )_2  \,
 \right > 
\label{C2S_rad}
}
where $w$ is the vertical component of the velocity,
$s_t$ the turbulent entropy fluctuation and 
$f_r(\xi_r,m) \equiv \left ({\partial \xi_r \over \partial r}\right )^2$, where
$\xi_r$ is the radial component of the eigenfunction, and  $g_r$ a function that involves
the first and second derivatives of $\xi_r$ (see Eq.\ (9) of \citet{Samadi02I}).
Quantities labelled with  1 and 2  denote two spatial and temporal positions, hence
$\left < w_1^2  w_2^2\right >$
and $\left < \left ( w s_t     \right)_1
\left ( w s_t  \right )_2  \,
 \right >$  correspond to two-point fourth-order correlation products. These correlation products are usually
approximated by expressions involving  second-order products only (closure model). In \cite{Samadi00I}, the simplest
approximation was used i.e the quasi-normal hypothesis. We study here consequences of using  a closure model closer to reality (i.e the
CMP from Paper I). Both are recalled in the next section.

\section{Closure models}

\subsection{The quasi-normal approximation}
\label{qna}
The  QNA  \citep[][Chap VII-2]{Lesieur97} is adopted in \citet{Samadi00I} as a convenient means 
to decompose the fourth-order velocity correlations in terms of a product of second-order vertical velocity correlations, that is, one uses
\begin{eqnarray}
\label{fourth-order}
 & & \langle w_1^2 w_2^2 \rangle_{QNA}  = 2 \, \langle w_1 w_2  \rangle^2 +  \langle w_1^2  \rangle \langle w_2^2 \rangle  \nonumber \\
 & & \left < \left ( w s_t \right)_1 \left ( w s_t  \right )_2  \, \right >_{QNA} = \left<w_1 w_2\right> \left< s_{t1} s_{t2} \right> \; ,
\end{eqnarray}
where $s_t$ is considered as a passive scalar.\\
This approximation (Eq.~(\ref{fourth-order})) remains strictly valid for normally distributed fluctuating quantities with zero mean.
As shown  by  \citet{Kraichnan57} in the context of turbulent flows and  \citet{Stein67} in the solar context,
the cumulant (the deviation from the QNA) can be large and therefore not negligible.
The CMP presented in Paper\ I was shown to be a significant improvement on the QNA for the one-point correlation products. 
However, we need two-point correlation products here (see Eqs.\ (\ref{C2R_rad}) and (\ref{C2S_rad})). A generalization of the CMP 
for two-point correlation products is therefore developed in Sect.~\ref{TPts} below. \\
The second-order correlation products in Eq.~(\ref{fourth-order}) are expressed in the Fourier domain $(\vec k,\omega)$ where $\vec k$ and $\omega$ are the wavenumber and the frequency associated with a turbulent element \citep[see][for details]{Samadi00I}.
%---------------------------------------------------------------------

\subsection{The closure model with plumes}
\label{TPts}
\begin{figure}[t]
\begin{center}
\includegraphics[height=7cm,width=9cm]{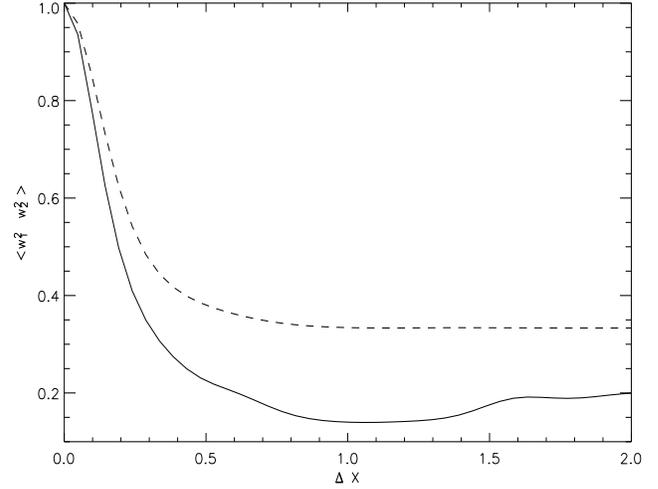} \\
\caption{ Fourth-order correlation function calculated in the quasi-adiabatic 
zone directly from the 3D numerical simulation (solid line) and using the QNA 
approximation (Eq.~(\ref{fourth-order}); dashed lines). The fourth-order moments are
 presented as a function of the correlation length ($\Delta X$), and the two curves are normalized 
 so as to emphasise only their $\bf{k}$ dependency. }
\label{correlation_1}
\end{center}
\end{figure}

The closure model with plumes (see Paper I) has been established only for one-point correlation products.
Here we generalize the CMP to two-point correlation products. 
We start in  Fig.~\ref{correlation_1} by comparing  the correlation product $<w^2_1 w^2_2>$ calculated  directly from  3D numerical simulations
 obtained from the Stein \& Nordlund code (see Sect. \ref{calculation_method}) with those calculated using  
Eq.~(\ref{fourth-order} of the QNA with second-order correlation products taken from the 3D simulation.
The question is whether the modelling of the $\vec k$ dependency on the two-point correlation function
 by the QNA can be used. For the sake of simplicity, we assume that the QNA can be used for the $\omega$ dependency.

The correlation products $\langle w_1^2 w_2^2 \rangle$ in Fig.~\ref{correlation_1}
are normalized  so as to compare only the $\vec k$ dependency of these quantities.
In the quasi-adiabatic region, the line width at half-maximum of the QNA and
the numerical product are roughly the same.
Discrepancies at high values of $\Delta X$ (the correlation length) are expected to have a negligible influence on the correlation product.
Hence, we assume that the modelling of the $\vec k$ dependency on the two-point correlation product by the QNA is valid due to
 a small difference between the line width at half-maximum. 
Hence it is legitimate to
 use the $( \vec k,\omega)$ dependency given by the QNA. One then needs only to correct
  the value of the correlation product at $( \vec k=\vec 0,\omega=0)$
  (which corresponds to the one point correlation function) with the CMP (see Paper I) for the turbulent Reynolds stress term 
  contribution. We use the interpolation formula of \cite{GH2002} for the FOM of the velocity (Paper I, Eq.~(13))
\begin{equation}
\label{4QNA}
<w^2_1 w^2_2>_{CMP} = (1+\frac{1}{3} S_{w}^2) <w^2_1 w^2_2>_{QNA} \; ,
\end{equation}
with $\langle w^2_1 w^2_2 \rangle_{QNA}$ given by Eq.~(\ref{fourth-order})
%\begin{equation}
%<w^2_1 w^2_2>_{QNA} = \Big(2<w_1 w_2>^2+<w^2_1><w^2_2>\Big) \,
%\end{equation}
 the skewness $S_w$ is calculated from the CMP (see Paper I for details). \\

\begin{figure}[t]
\begin{center}
\includegraphics[height=7cm,width=9cm]{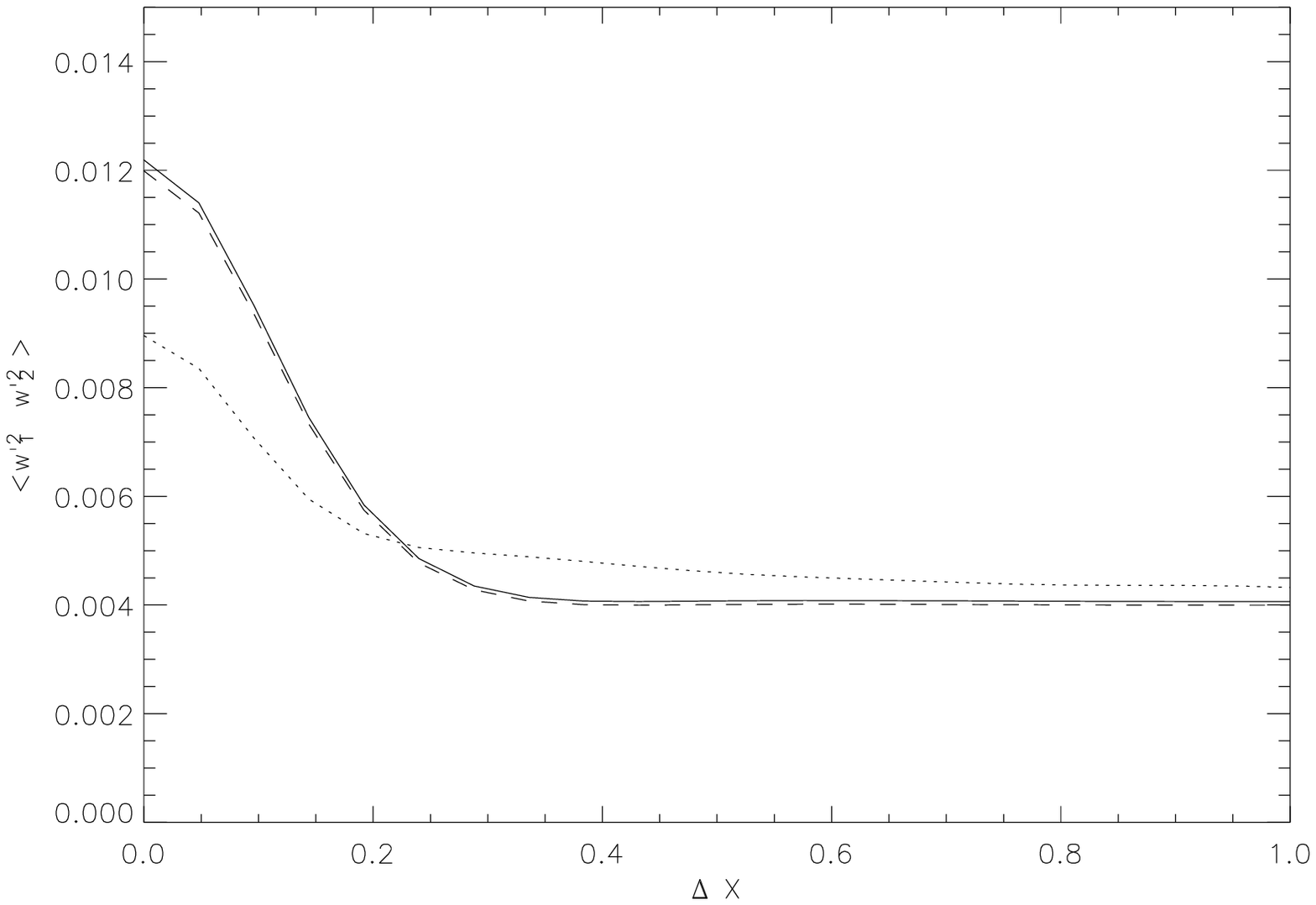}
\includegraphics[height=7cm,width=9cm]{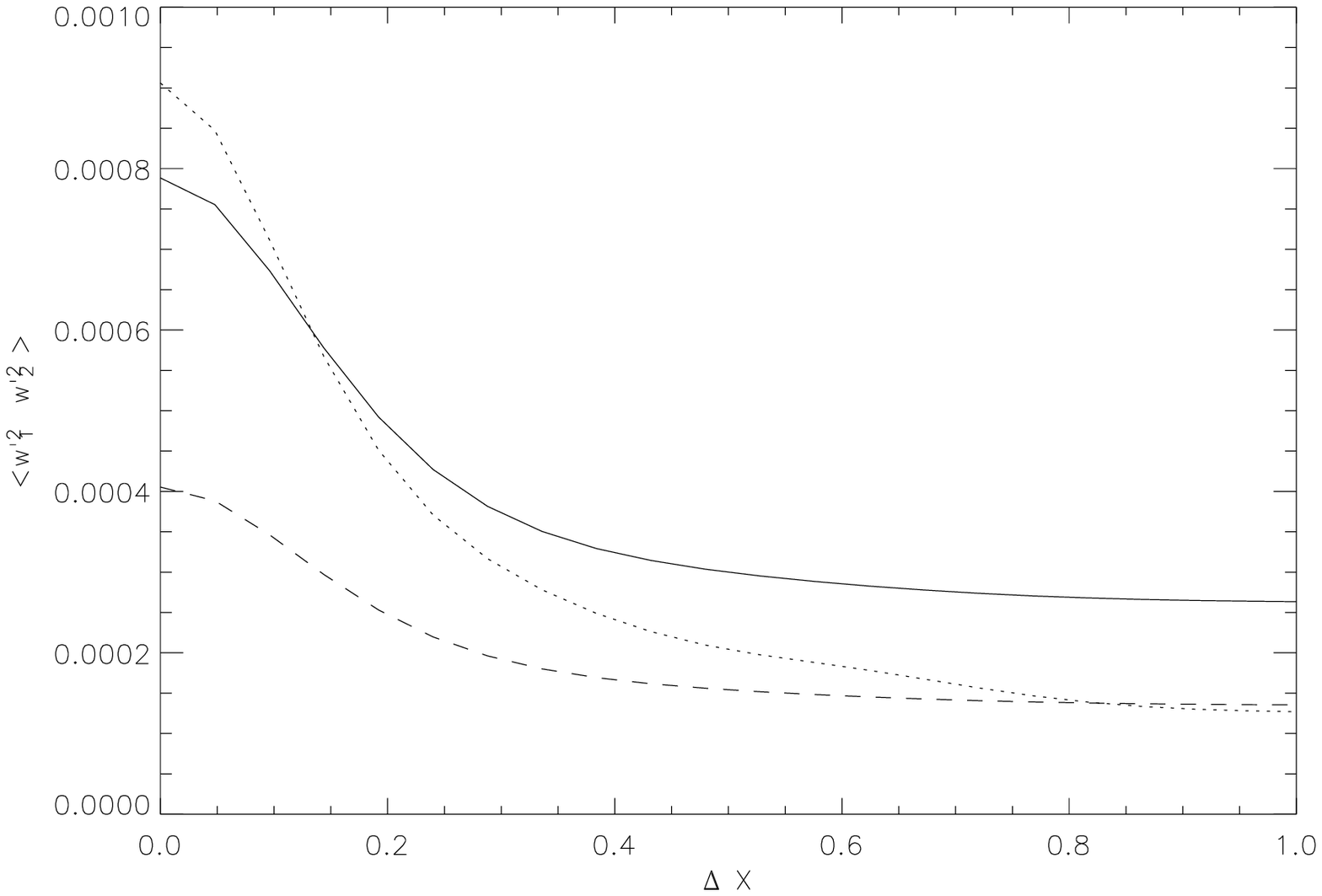} \\
\caption{Fourth-order correlation function calculated in the superadiabatic zone
(at the top) and in the quasi-adiabatic zone (at the bottom) directly from the 3D numerical 
simulation (dotted line), using the QNA approximation (Eq.~\ref{fourth-order}; dashed lines) and 
using the CMP (Eq.~\ref{4QNA}; solid line).}
\label{correlation_2}
\end{center}
\end{figure}

In Fig.~\ref{correlation_2}, calculations using  Eqs.~(\ref{fourth-order}) and~(\ref{4QNA}) are compared to the direct numerical correlation product.
 The above  generalization of  the CMP to two-point correlation products
 provides a good approximation mainly in the quasi-adiabatic region where the CMP is the more accurate one (see Paper I).
  The $\bf{k}$ dependence is approximatively modelled by the QNA (Fig.~\ref{correlation_1})
  except for large correlation lengths ($\Delta X > 0.2 \, Mm$), but these contribute only negligibly
   to $<w_1^2 w_2^2>$. However, in the superadiabatic zone, the generalization of the CMP and the QNA
   both fail to describe the two-point correlation function. In that zone,
   the temperature gradient is varying quickly, which is not the case in the CMP. In the plume model (Paper I) the temperature 
gradient appears only through a polytropic law, and
   for sake of simplicity we assume an isentropic atmosphere. In addition, for modelling the FOM $<w^4>$,  
the interpolated formula derived by \cite{GH2002} (Paper~I, Eq.~(13)) is not valid in the superadiabatic zone.
 Thus, in this zone the treatment of Eqs.~(\ref{fourth-order}) and~(\ref{4QNA}) will introduce an energy excess injected
into high-frequency $p$~modes.\\


\section{Calculation of the theoretical $p$~mode excitation rates}
\label{calculation_method}

The rate ($P$) at which energy is injected per unit time into a mode is
calculated according to the set of Eqs.~(\ref{C2R_rad}) - (\ref{fourth-order}) 
when the QNA is used and Eqs.~(\ref{C2R_rad})-(\ref{4QNA}) using the CMP (see Sect.~\ref{cal_p}).
 The calculation thus requires the knowledge of four different types of quantities:
\begin{itemize}
\vspace{-0.0cm} \item[1)] Quantities that are related to the
oscillation modes: the eigenfunctions ($\xi_r$) and associated
eigenfrequencies ($\omega_0$).
\vspace{-0.0cm} \item[2)] Quantities that are related to the spatial
and time-averaged properties of the medium: the density $\rho_0$, the vertical velocity $\tilde{w}$, the entropy $\tilde s$, and  $\alpha_s=\partial P_0 / \partial \tilde{s}$.
\vspace{-0.0cm} \ \item[3)] Quantities that contain information about
spatial and temporal correlations of the convective fluctuations:
$E(k)$, $E_s(k)$, and $\chi_k(\omega)$.
\vspace{-0.0cm} \ \item[4)]  Quantities that take anisotropies into account:
$a$ and $\Phi$. The value of $a$ is the mean horizontal fractional area of the updrafts (see Paper\ I), whereas
 $\Phi$
  measures the anisotropy of turbulence and is defined according to \cite{Gough77} \citep[see also][for details]{Samadi00I}
   as:
\begin{equation}
\Phi=\frac{<w^2>}{<u^2>} \; ,
\end{equation}
where $u^2=w^2+u_h^2$ and $u_h$ is the horizontal velocity. 
\end{itemize}
Both $a$ and $\Phi$ are necessary to describe the flow because $a$ measures the geometric anisotropy between
up and downflows while $\Phi$ corresponds to the measure of the velocity anisotropies. However, these two quantities are linked 
because of mass conservation. An explicit relation can be easily derived between them using the 
formalism developed in Paper~I to obtain
\begin{equation}
\Phi = \frac{a(1-a)\delta w^2 + a <\tilde{w}^2>_u + (1-a)<\tilde{w}^2>_d}{a(1-a)\delta w^2 + a <\tilde{u}^2>_u +(1-a)<\tilde{u}^2>_d}
\end{equation}
where the $~\tilde{ }~$ refers to the velocities of only one flow (updraft or downdraft) and $\delta w$ is defined as in Paper~I. 
For consistency reason, $a$ and $\Phi$ are provided by the 3D numerical simulation. 

\subsection{The solar case}
\label{cal_p}

Calculations of the eigenfrequencies and eigenfunctions (in point 1) above)
 are performed as in \citet{Samadi02I} on the basis of a 1D solar model built
  according to Gough's (1977)\nocite{Gough77} non-local formulation of  the mixing-length theory (GMLT hereafter).

 The spatial and time-averaged quantities in point 2) are  obtained  from a 3D
simulation of the solar surface.
The  3D simulations used in this work were built with Stein \& Nordlund's 3D numerical
code \citep[see][]{Stein98,Samadi02II}. Two simulations with different spatial mesh grids are
considered, namely 253$\times$253$\times$163 and 125$\times$125$\times$82, in order to verify that the results are not sensitive to the
spatial mesh resolution.

Finally, for the
quantities in point 3) the total kinetic energy
contained in the turbulent kinetic spectrum, $E(k)$, its depth
dependence, and its $k$-dependence are obtained directly from a 3D
simulation of the uppermost part of the solar convective zone.
It was found in \citet{Samadi02II} from 3D simulations that a Gaussian --\, usually used
for modelling $\chi_k$ \,-- is inadequate: a Lorentzian
fits the frequency dependence of $\chi_k$ best.  Hence, we adopt a Lorentzian here for $\chi_k$.\\

\subsection{Calculation of the power injected into the solar $p$~modes with the CMP}
\label{cal_p2}
\begin{figure}[t]
\begin{center}
\includegraphics[height=7cm,width=9cm]{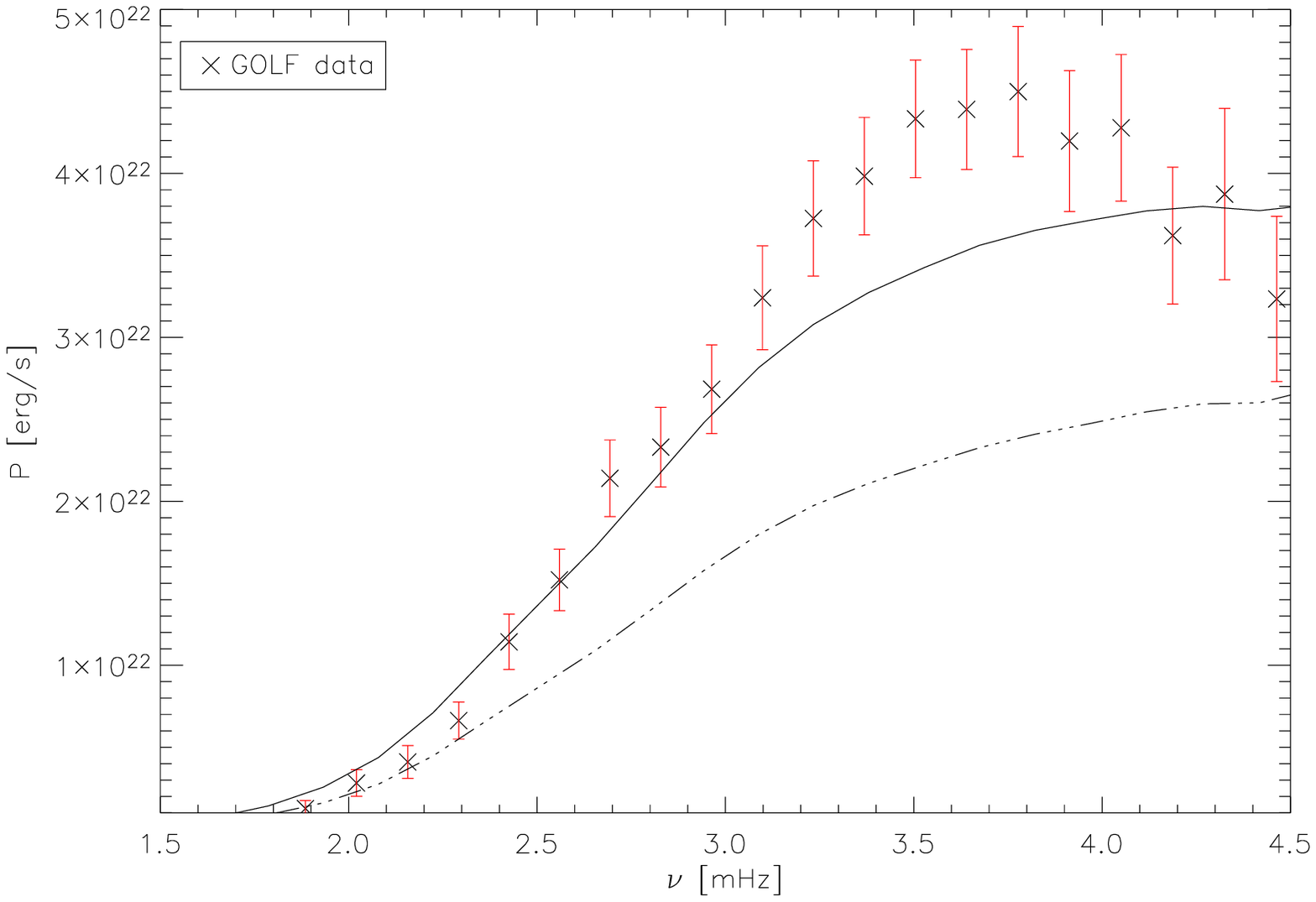} \\
\includegraphics[height=7cm,width=9cm]{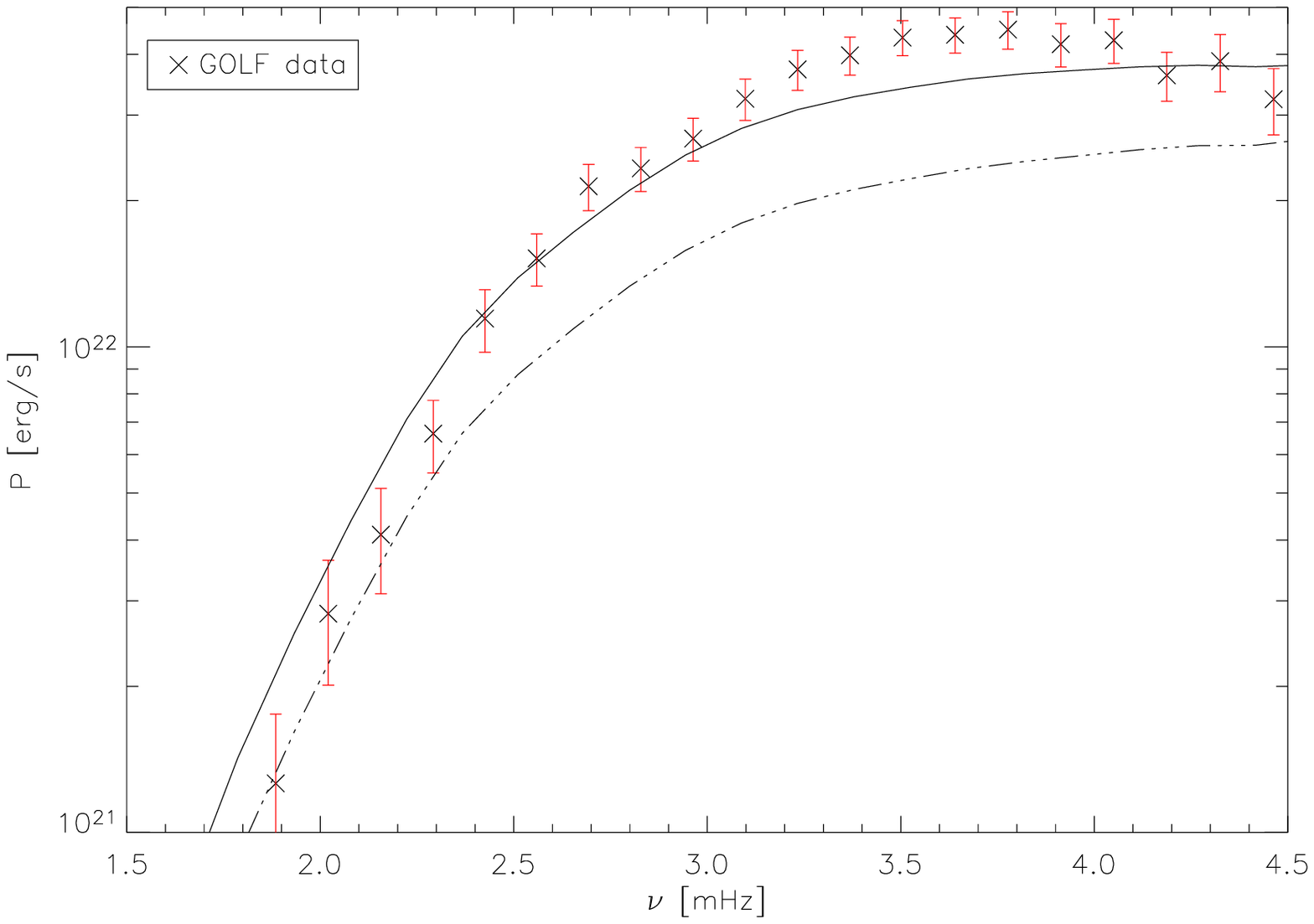} \\
\caption{Rate $P$ at which acoustic energy is injected into the solar radial modes.
\emph{Only the Reynolds stress contribution is computed}.
 Cross dots represent $P$ computed from \cite{Baudin05}
 solar seismic data from the GOLF instrument (see Sect.~\ref{OBDS}).
   The associated error bars take into account uncertainties both 
    from the line width ($\Gamma_\nu$) and from the maximum height
    of the mode profile ($H$).
    The curves represent theoretical values of $P$ computed
    as explained in Sect.~\ref{calculation_method}: dash-dotted lines correspond
    to the calculation of $P$ using the QNA closure model, and 
    solid lines represent $P$ using the CMP for the Reynolds stress term.
    We present the results in linear (at the top) and logarithmic scale (at the bottom).}
\label{Reynolds}
\end{center}
\end{figure}
We use the generalized CMP for two-point correlation functions
presented in Sect.~\ref{TPts} (Eq.~(\ref{4QNA})) to model
 the Reynolds-stress source term.
By replacing Eq.~(\ref{fourth-order}) with Eq.~(\ref{4QNA}) in Eq.~(\ref{C2R_rad}),
 the calculation of $C_R^2$ \citep[as in][]{Samadi00I} yields:
\begin{eqnarray}
\label{C2R_3bis2}
C_R^2  &=&   {64 \over 15} \, \pi^{3} \, \int_{0}^{M} dm  \, (1+\frac{1}{3}S_{w}^2) \,
\rho_0 \left (\deriv { \xi_r} {r} \right )^2
\int_0^\infty dk \, \nonumber \\
&\times& \int_{-\infty}^\infty d\omega  \, \frac {E^2(k)} { k^4}  \chi_k ( \omega_0 + \omega, r) \, \chi_k (\omega, r)  \; .
\end{eqnarray}
Equation~(\ref{C2R_3bis2}) shows that the CMP causes an increase in the power injected into $p$~modes
in comparison with calculation using only the QNA.
\begin{figure}[t]
\begin{center}
\includegraphics[height=7cm,width=9cm]{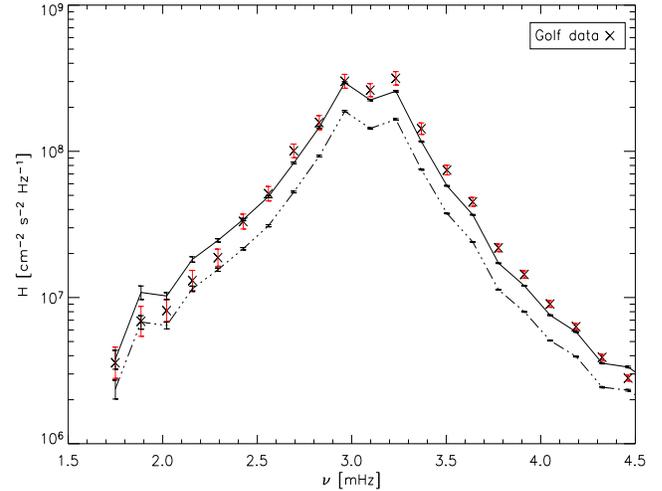} \\
\caption{Mode height $H$ calculated as explained in Sect. \ref{OBDS}
 \emph{using only the Reynolds stress contribution}.
The solid  (resp. dash-dotted) line represents $H$ calculated with the CMP (resp. QNA) closure model,
and cross-dots represent GOLF data with associated error bars.
 Error bars associated with the curves are due to  mode line widths that are taken
 from observations (see Eq.~(\ref{def_H})).}
\label{H}
\end{center}
\end{figure}
On the other hand, the entropy source term, $C_S^2$, is still  computed
using the QNA closure model \citep[see][for details]{Samadi00I}.\\
%%%%%%%%%%%%%%%%%%%%%%%%%%%%%%%%%%%%%%%%%%%%%%%%%%%%%%%%%%%%%%%%%%%%%%%%%%%%%%%%%%%%%%%%%%%%%%%%%%%%%%%%%%%%%%%%%%%%%%%%%
\section{Observational data and inferring observed excitation rates}
\label{OBDS}
The observational data set selected here for comparison with theoretical calculations
was obtained with   the GOLF instrument,
onboard SOHO. GOLF \citep{Gab97} is a spectrometer measuring velocities
of the photosphere integrated over the whole solar disc. Its location on the
space platform yields a very good signal-to-noise ratio and also continuous
observations (the actual duty cycle reaches almost 100\%). This latter
characteristic greatly improves the signal to noise ratio in the Fourier
spectrum.\\
However, GOLF suffers from some technical problems, which restricts the
measurements to one wing of the Na~D$_1$ line instead of both wings. This
results in a more difficult absolute calibration of the measured velocity and
thus a possible bias (which does not exceed 20\% in terms of the acoustic rate of
excitation). Characteristics of the data set used here are described in \citet{Baudin05}.\\
 These observations correspond to two periods when GOLF was observing in
the same instrumental configuration (blue wing of the Na line) with a duration
of 805 and 668 days, starting on April 11, 1996
and November 11, 2002, respectively. The level of solar activity was different during
these two periods, but the measured excitation rate shows no dependence on
activity,
as the increase in width compensates for the decrease in height of the peaks,
as shown by \cite{Chaplin00} or \cite{Jim03}.\\
The GOLF results were compared to BiSON observations and are
compatible with them over a wide frequency range. A discrepancy appears at high
frequency ($\nu >$  3.2 mHz). As the height and width of peaks in the Fourier
spectrum are affected by the presence of noise and gaps in the data (see Chaplin
et al 2003), GOLF was chosen for the comparison model/observations.
We consider only the $\ell=1$ modes for which their properties (line-width, amplitude)
 are more accurately determined than the $\ell=0$ modes \citep[see][for details]{Baudin05}.

In order to compare theoretical results and observational data,
the mode excitation rates are inferred from the observations
according to the relation
\begin{equation}
\label{Pobs}
P_{\rm obs}(\omega_0) = 2 \, \pi \,
\Gamma_\nu \, {\cal M} \, v_s^2 (\omega_0)
\end{equation}
where ${\cal M} \equiv I / \xi_{\rm r}^2(h) $ is the mode mass, $h$
the height above the photosphere where oscillations are measured,
$\Gamma_\nu = \eta/ \pi $ the mode linewidth at half maximum (in Hz),
 and $v_s^2$ the mean square of the mode surface velocity. The last 
is derived from the observations according to 
\begin{equation}
\label{vs}
v_s^2 =  \pi  \,  H \, \Gamma_\nu  \, C_{\rm obs}
\end{equation}
where $H$ is the maximum height of the mode profile in the power
spectrum and $C_{\rm obs}$ the multiplicative constant factor that 
depends on the observation technique \citep[see][]{Baudin05}. 
Equation~(\ref{vs}) supposes that the mode line profiles are symmetric, but it is well known that the mode profile deviates from a Lorentzian. 
However, \citet{Baudin05} show that this equation is accurate enough for the evaluation of
the mean square of the mode velocity, Eq.~(\ref{vs}). 
On the other hand, the mode asymmetry
is taken into account when determining mode line widths from observational data.

The mode mass is very sensitive to altitude at high frequency
(see Fig.~1 of Baudin et al 2005), so the layer
($h$) where the mode mass is evaluated must be properly estimated
to derive correct values of the excitation rates.  Indeed, solar
seismic observations in Doppler velocity are usually measured
from a given spectral line. The layer where oscillations are measured then
depends on the height where the line is formed. 
%Different
%instruments use different solar lines and then probe a different
%region of the atmosphere.  The BiSON instruments use the KI line whose
%height of formation is estimated at the height $h \approx 280$\ km
%while
 The GOLF instrument uses the Na I D~1 and D~2 lines whose height
of formation is estimated at the height $h \approx 340$\ km
\citep[see][]{Baudin05}.

% the difficulty to determine the height in the solar atmosphere where
%observations are performed. This depends on the formation
%µheight of the atomic line used by the instrument (see Baudin et al 2005)
%through the mode mass, which .

% From the same reference, we have chosen
%here the altitude of 340~km for an estimation of Na~D$_1$ line formation.
%  In Fig.~\ref{Reynolds} and \ref{CR+CS}, $P_{\rm obs}$ is
%calculated with mode masses evaluated at $h= 280$\ km for the BiSON
%data set  and at $h=340$\ km for the GOLF data . Both data sets are
%from \citet{Baudin05}.

As an alternative to comparing theoretical results and observational data,  \citet{Gough05}
propose to  derive the maximum height  of the mode profile
($H$) from the  theoretical excitation rates and the observed mode
line width according to the relation:
\begin{equation}
\label{def_H}
H =  \frac{P}{2\pi^2 {\cal M} \Gamma_\nu^2 C_{\rm obs} } \, ,
\end{equation}
where $C_{obs} = 2.59$ for $\ell=1$ modes.\\
Representation of the excitation rates themselves (Eq.~(\ref{Pobs})) emphasises
 disagreement at high frequencies, whereas disagreement
at low frequency is more apparent  with a representation using the profile height (Eq.~(\ref{def_H})).
Note that in the case of the observable height, only the  slopes are the meaningful quantities, as the amplitude magnitude 
depends on the phase of  the solar cycle when the observations were recorded.  

As the maximum height $H$ strongly depends on the  observation technique,
one cannot compare values of $H$
 coming from two different instruments. In Fig.~\ref{H}, we therefore
 plot the product $HC_{\rm obs}$, a quantity that is less dependent on the observational data (but still through $\cal M$).
Note that for ease of notation, $HC_{\rm obs}$ is noted $H$ in the following.

\begin{figure}[t]
\begin{center}
\includegraphics[height=7cm,width=9cm]{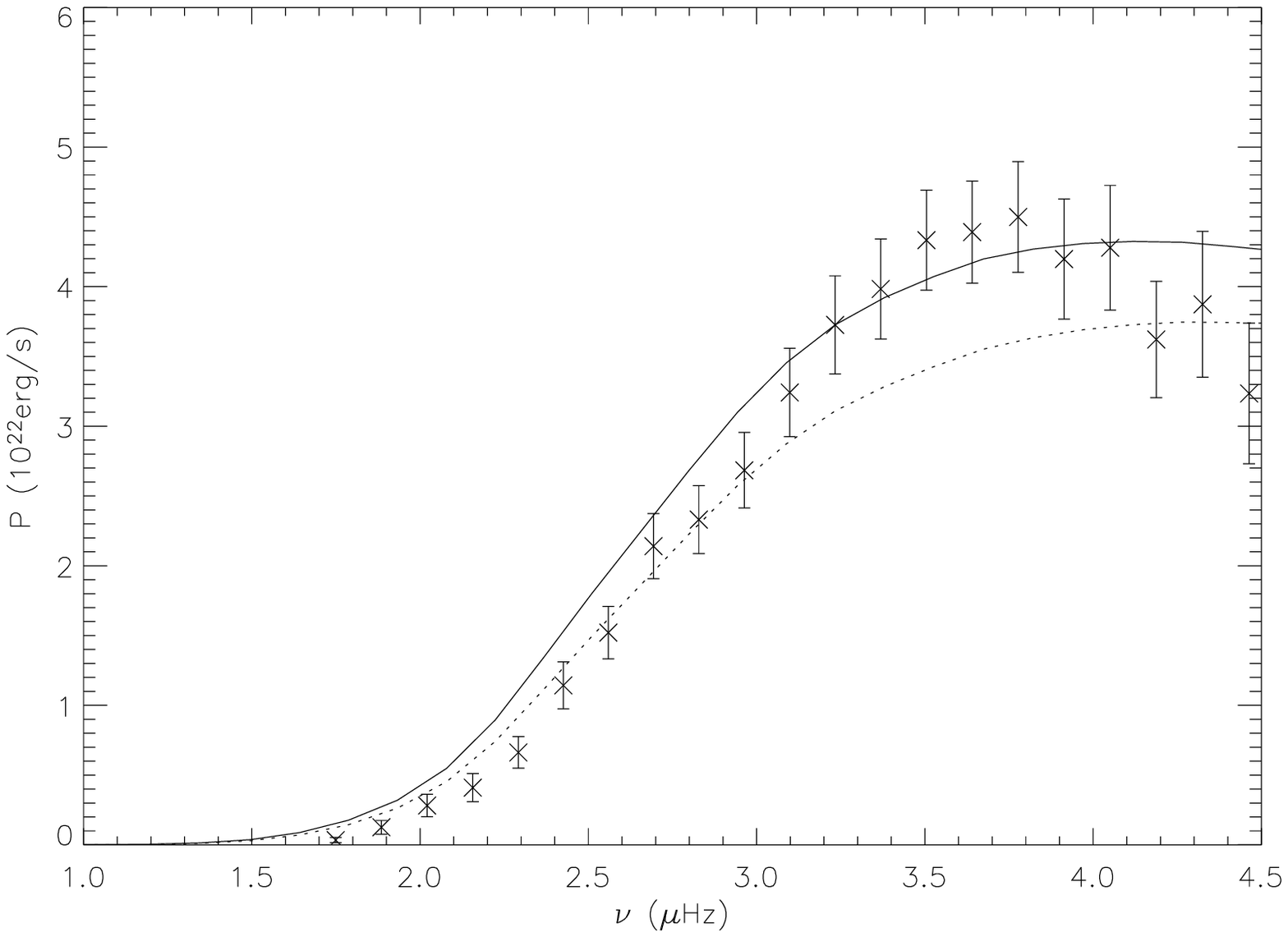} \\
\includegraphics[height=7cm,width=9cm]{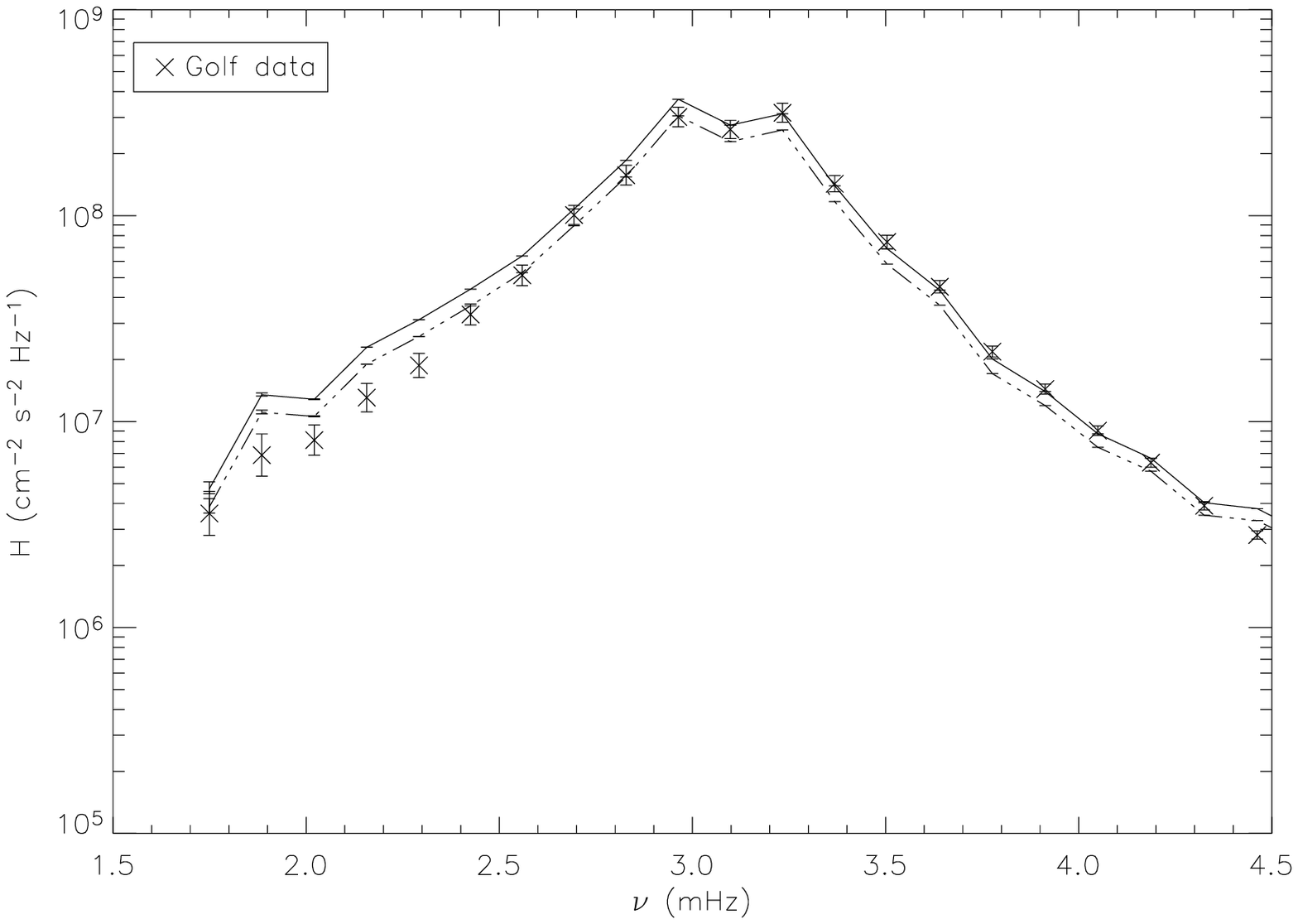} \\
\caption{ {\it Top:} Rate ($P$) at which acoustic energy is injected into the solar radial modes as a function of frequency.
 Cross dots represent $P$ computed from the \cite{Baudin05}
 solar seismic data from the GOLF instrument (see Sect.~\ref{OBDS}).
    The curves represent theoretical values of $P$  computed
    as explained in Sect.~\ref{calculation_method}: 
    the solid line represents $P$ using both the Reynolds stress (using the  CMP) and entropy  source contributions. 
    The dotted line corresponds to the calculation for the Reynolds stress term using only the CMP closure.
    {\it Bottom:}Mode height ($H$) calculated as explained in Sect.~\ref{OBDS}.
 The solid line represents $H$ calculated with the CMP closure model, \emph{using the Reynolds stress and entropy fluctuation contributions}.
 The dotted line represents $H$ computed with the CMP closure model, \emph{using only the Reynolds stress contribution}. 
 Cross-dots represent GOLF data with the associated error bars.
 Error bars associated with the curves are due to  mode line widths 
 that are taken from observations (see Eq.~(\ref{def_H})). Only observations near minimum solar activity have been used, and they correspond to the second 
period as explained in Sect.~\ref{OBDS}. } 
\label{CR+CS}
\end{center}
\end{figure}

 It is important to  stress that the mode height ($H$) calculated from the  theoretical excitation rates (Eq.~(\ref{Pobs}))
 depends on the observations through the line width $\Gamma_\nu$.
This is why in Figs.~\ref{CR+CS} and \ref{H} error bars appear in the theoretical results. 
In any case, the observational data can be characterised by
at least three main features that the  theoretical calculations
(see above)  must reproduce:
\begin{enumerate}
\item the frequency dependence from low to medium frequencies ($\nu < 3$ mHz).
\item the maximum of amplitude at $3$ mHz for $H$ and the slope for frequencies between $3$ and $4$ mHz
or  a  nearly flat maximum between $\nu \approx 3.8$ mHz and $4$ mHz for $P$.
\item the slope at very high frequencies $\nu > 4$ mHz.
\end{enumerate}

\section{Comparison between theoretical and observed excitation rates}
\subsection{Turbulent Reynolds stress contribution}
\label{Reyn_only}

Figure~\ref{Reynolds} compares  the observed power $P$ injected into solar $p$~modes with the theoretical one computed
with only the turbulent Reynolds stress term assuming either the CMP or the QNA closure models.  Figure~4 shows the
associated heights $H$ as computed according to Eq.~(13). 
The comparison shows that  the closure model  has a significant effect on the resulting  excitation rates.
Indeed, the CMP induces an increase in the energy injected into the mode by about a factor two in comparison
with the QNA closure model and brings the theoretical excitation rates closer  to the observational ones.
This energy increase is not uniform in terms of frequencies, 
due to the variation in the skewness
 with the depth ($z$) (see Paper I for details) and
 to the fact that the mean square velocity amplitudes
 of the turbulent elements decrease with depth. 
 Indeed, at the top of the convection zone where the highest frequency modes are confined,  the
 inefficiency of the convective transport causes  an increase in the velocities. 
 Thus the effect of the flow anisotropy becomes dominant for such high-frequency modes.\\
 \emph{At low frequencies ($\nu < 2.5 \, $ mHz)},
  the turbulent Reynolds stress contribution reproduces the observed power $P$  (Fig.~\ref{Reynolds})
   within the observational  uncertainties. As best emphasised in Fig.~4, 
   it is possible that the theoretical results are slightly
  overestimated, although this remains within the observational error bars.\\
\emph{At intermediate frequencies $4 > \nu >  3  ~mHz$)},
the turbulent Reynolds stress term  is  not sufficient to reproduce the observations, so 
the  additional excitation coming from  entropy fluctuations is necessary.\\
\emph{At high frequencies $\nu >  4  ~mHz$)},
Observational data seem to indicate a decrease in the power, which is not reproduced by the theoretical power.

\subsection{Adding the entropy fluctuation contribution}
\label{both}

To proceed further, we add the $C_S^2$ contribution (Eq.~(\ref{C2S_rad})). Results for the excitation rate 
and the maximum height  are presented
 in Figs.~\ref{CR+CS} and~\ref{H}, respectively. 
The additional (positive) entropy contribution causes an overall increase in the excitation rates as shown in
Fig.~\ref{CR+CS}.  
 The theoretical modelling now reproduces the maximum of the power supplied to the modes when compared with the observational data.
For the frequency behaviour of the excitation rate and height, (Figs.~\ref{CR+CS} and \ref{H}) show:

 {\it At low  frequency }($\nu \in [1.6 \, {\rm mHz}; 3 \, {\rm mHz}]$). 
 We pointed out in Sect.\ref{Reyn_only} that the contribution from the Reynolds stress term
 can be sufficient for reproducing the GOLF data, perhaps even overestimating it. 
 The combination of both Reynolds stress and
 entropy fluctuation  is too large compared with the observation, and the resulting slope differs 
 from the observational one in this frequency domain. Note however, that in Fig.~\ref{CR+CS} error bars represent  
 1 $\sigma$ error bars.
(Fig.~\ref{CR+CS}).

 {\it For intermediate  and  high  frequencies} ($\nu  \in [ 3 \, ; 4 \,] ~ {\rm mHz}$), 
 the Reynolds (CMP) and entropy  excitation model
reproduces the $\nu$ variation in $P$. 
This is confirmed with the $H$ representation (Fig.~\ref{CR+CS} at the bottom).
However from a theoretical point of view, 
 the description of the behaviour at high frequencies ($\nu > 4 ~ {\rm mHz}$) is more complicated because these $p$~modes are
mainly excited in the superadiabatic zone, which is difficult to model properly. 
On the observational side, it must be kept  in mind 
that  even data with a signal-to-noise ratio as good as GOLF lead to linewidths 
difficult to measure at high frequencies.

\section{Discussions and conclusions}
\label{conclusion}

 We use a closure model  (CMP, Paper I) that is more realistic than 
 the usual QNA approximation to model the correlation  products in a semi-analytical 
 description of the excitation process of solar $p$~modes.
 The present excitation  model gives the theoretical \emph{slope} of the power at intermediate and high frequencies ($\nu \in [2.5 \,
 mHz; 4 \, mHz]$), which agrees with  the observed data.
  We also find that  including the CMP  causes a global increase in the injected power. 
  This brings the power computed with  the Reynolds stress
  contribution alone closer to (although, at intermediate frequency, still below) the observations. 
  On the other hand, the power obtained by including both the Reynolds stress and the entropy fluctuation
  contributions reproduces the observations at the maximum of the excitation rates.
  The comparison can now be made in linear scale, hence at lower frequencies there is still a small over-estimation 
  (which amounts roughly to a few per cents 
  and the errors bars represent 1$\sigma$ error bars).
%  is  too large compared with the observations (note however that the comparison can now be made
 % in linear scale and that  the overestimation amounts roughly to  $25\%$  
%  and the errors bars are 1$\sigma$ error bars)
 The reason for this overestimation  cannot be attributed to the CMP. 
 Indeed,  the Reynolds stress contribution was  
compared to the 3D numerical simulation (see Paper~I), and the one-point fourth-order 
moment $<w^4>$ was found to agree with the simulation result.
The remaining departure from the numerical simulation shows that the CMP actually 
underestimates the FOM in the
quasi-adiabatic region, so correcting for this bias would result
in an even larger overestimation of the power.

 Various sources of discrepancies are likely to exist: the separation of scales used in the formalism
 that consists in assuming that the stratification and the oscillations have characteristic scale lengths larger
 than the eddies contributing to the excitation \citep[see][for details]{Samadi00I}. 
The physical description of the outer layers in the 1D solar model can also play an important role
directly through the velocity and indirectly through  the eigenfunctions.
In this paper, we use Gough's (1977) \nocite{Gough77} non-local
formulation of the mixing-lenght theory
which shows  an improvement in comparison with the local formulations
in terms of the maximum of power $P$ \citep{Samadi05b} by about a few percent. 
Concerning the excitation model itself, some improvements in the modelling of Reynolds and entropy contributions
that ought to be  investigated are outlined below.

\subsection{Turbulent Reynolds stress tensor contribution shortages}

\begin{figure}[t]
\begin{center}
\includegraphics[height=7cm,width=9cm]{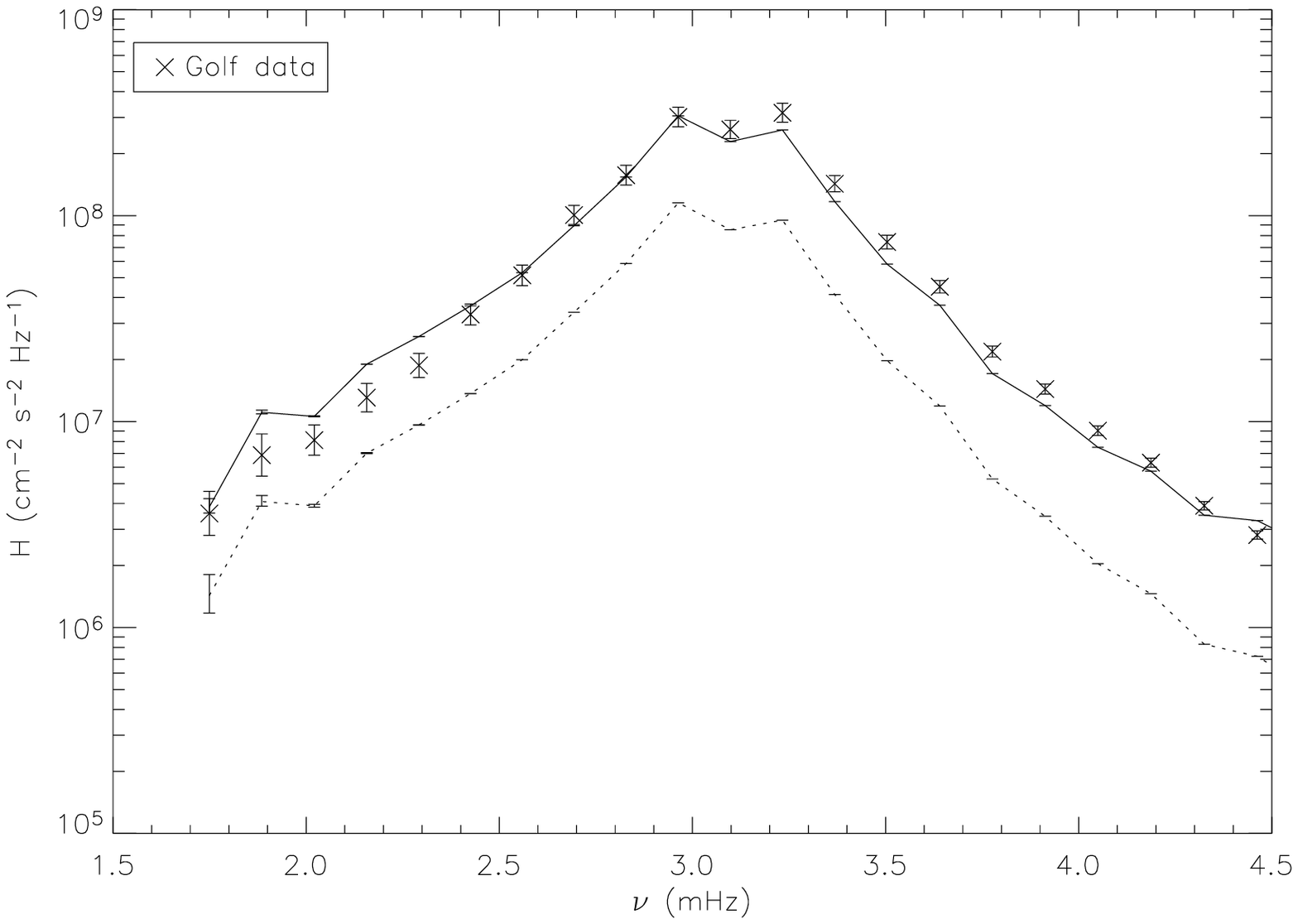} \\
\caption{Mode height $H$ calculated as explained in Sect. \ref{OBDS}
 \emph{using only the Reynolds stress contribution}.
 Solid lines represent $H$ calculated with the CMP closure model and 
 dots-line is the same except that a Gaussian is used for $\chi_k$. 
 Crosses represent GOLF data with associated error bars.
 Error bars associated with the curves are due to  mode line widths which are taken from observation (see Eq.~(\ref{def_H})).}
\label{H}
\end{center}
\end{figure}

At low frequencies,  a possibly  small overestimation of the Reynolds stress contribution  can be attributed to 
the frequency dependent factor ($\chi_k$, see Eq.~10 in Sect.~\ref{cal_p}). 
\cite{Gough05} use a Gaussian $\chi_k$
whereas \cite{Samadi02I} use a Lorentzian factor. 
In Fig.~\ref{H}, we present the calculation assuming a Gaussian and a Lorentzian for $\chi_k$.            
As shown there, the frequency-dependent factor $\chi_k$ is likely between these two regimes. 
In the quasi-adiabatic convection zone, plumes are well-formed, 
and the convective system must
be treated as composed of two flows (see Paper~I). 
Hence, the upflows that are less turbulent can be modelled by a white noise (Gaussian),
 but downflows are turbulent creating a departure from a Gaussian. We expect this effect to cause 
a decrease in the theoretical power    
and bring it closer to the observation. A rough idea 
can be obtained by taking this effect into account as follows:  we split the  computation of the  
power supplied into the modes into two parts. Those parts correspond to 
upflow ($\chi_k$: Gaussian) and to downflows ($\chi_k$: Lorentzian) . 
The result indicates a decrease in the power at low
frequency, which brings the theoretical power closer to the observation.
This is true mainly for low-frequency modes, which are less sensitive to the superadiabatic zone where plumes are formed,
 because this region cannot be modelled by such a simple model. This issue needs further investigation.

\subsection{Entropy source contribution shortages}

In the present model, the turbulent entropy fluctuations are assumed to behave as a passive scalar,
 in other words, the entropy fluctuations are assumed to be 
advected by the turbulent velocity field without dissipation. 
It means that the entropy field does not have any effect on the velocity field. 

This assumption 
associated with the QNA has the advantage of simplifying the closure of the 
fourth-order moments involving the entropy fluctuations (see Eq.~(\ref{qna})). However the biases 
introduced by this assumption remain to be evaluated. If the biases turn out to 
be large,  alternative models must be developed.\\

%Entropy source contribution at low frequencies.
% We suspect that this discrepancy arises because the entropy contribution is not properly modelled in the superadiabtic region where
% its effect is strongly concentrated and this affects all modes including those with low frequency. \\

%the CMP closure model is not adapted because it corresponds to $p$-modes
%which are mainly excited in the upper most layer of the convection region where the plume model fails because
%it does not take into account radiative cooling.
%However, here again,  the modelling of the entropy source term over-estimates the power injected into $p$-modes.\\

\subsection{Perspectives}

Finally, we stress that there is an additional dependency,
 the coefficient $a$, which is the mean fractional area of updraft on the horizontal plane (see Eq.9). It is a measure of the
asymmetry of the flows and a small variation in its value plays a major role on the excitation rates. 
This parameter has been fixed here
using the results of  3-D simulations. 
The influence of parameter $a$ is very important, as  a small variation of its value leads to an increase in 
 power $P$ through the skewness $S_w$ (see Paper~I). It is beyond the scope of this paper to estimate
 the true effect of a variation in this parameter because its value is linked to the physical properties of
 the flows through, for instance,  conservation of the mass flux.
  Hence a consistent approach is to investigate a set of different numerical simulations.\\
 The CMP closure model, indeed, strongly depends on the structure of the upper convection zone,
 which again emphasises that the structure of this region  is very important in the theoretical prediction of
  the power injected into
  the $p$~modes, because the skew introduced by the asymmetry increases with the departure of $a$ from the value $0.5$.
  It is then possible to obtain physical constraints on the asymmetry of the convection zone flows.\\
To understand what can affect $a$ is therefore an important issue, and in near future it will be necessary to study the
variation in $a$ with the type of star and from a hydrodynamical point of view to determine what the main
processes that are responsible for this asymmetry. One interesting issue is the influence of a magnetic
field on this parameter: as described by \cite{Weiss02} and \cite{Voegler05}, the effect of a strong magnetic field
induces a reduction in the typical length scale of convection, as well as the structure of the flows (hence the value of $a$).\\
The study of the mean fractional area $a$ as a function of the
magnetic field intensity  therefore represents  an interesting perspective for characterising
 $B$ from the excitation rates, at least  for stars
with an expectedly strong magnetic field.  \\

\begin{acknowledgements}
We are indebted to J.~Leibacher for his careful reading of the manuscript and his helpful remarks.
We thank {\AA} Nordlund and R.~F. Stein for making their code 
available to us.  Their
code was made at the National Center for Supercomputer 
Applications and
Michigan State University and supported by grants from NASA and NSF.
\end{acknowledgements}

%\bibliographystyle{../style/aa}
%\bibliography{bib}

\begin{thebibliography}{26}
\expandafter\ifx\csname natexlab\endcsname\relax\def\natexlab#1{#1}\fi

\bibitem[{{Baudin} {et~al.}(2005){Baudin}, {Samadi}, {Goupil}, {Appourchaux},
  {Barban}, {Boumier}, {Chaplin}, \& {Gouttebroze}}]{Baudin05}
{Baudin}, F., {Samadi}, R., {Goupil}, M.-J., {et~al.} 2005, \aap, 433, 349

\bibitem[{{Belkacem} {et~al.}(2006){Belkacem}, {Samadi}, {Goupil}, \&
  {Kupka}}]{Belkacem06a}
{Belkacem}, K., {Samadi}, R., {Goupil}, M.~., \& {Kupka}, F. 2006, \aap (in
  press)

\bibitem[{{Chaplin} {et~al.}(1998){Chaplin}, {Elsworth}, {Isaak}, {Lines},
  {McLeod}, {Miller}, \& {New}}]{Chaplin98}
{Chaplin}, W.~J., {Elsworth}, Y., {Isaak}, G.~R., {et~al.} 1998, \mnras, 298,
  L7

\bibitem[{{Chaplin} {et~al.}(2000){Chaplin}, {Elsworth}, {Isaak}, {Miller}, \&
  {New}}]{Chaplin00}
{Chaplin}, W.~J., {Elsworth}, Y., {Isaak}, G.~R., {Miller}, B.~A., \& {New}, R.
  2000, \mnras, 313, 32

\bibitem[{{Chaplin} {et~al.}(2005){Chaplin}, {Houdek}, {Elsworth}, {Gough},
  {Isaak}, \& {New}}]{Gough05}
{Chaplin}, W.~J., {Houdek}, G., {Elsworth}, Y., {et~al.} 2005, \mnras, 360, 859

\bibitem[{{Gabriel} {et~al.}(1997){Gabriel}, {Charra}, {Grec}, {Robillot},
  {Roca Cort{\'e}s}, {Turck-Chi{\`e}ze}, {Ulrich}, {Basu}, {Baudin},
  {Bertello}, {Boumier}, {Charra}, {Christensen-Dalsgaard}, {Decaudin},
  {Dzitko}, {Foglizzo}, {Fossat}, {Garc{\'{\i}}a}, {Herreros}, {Lazrek},
  {Pall{\'e}}, {P{\'e}trou}, {Renaud}, \& {R{\'e}gulo}}]{Gab97}
{Gabriel}, A.~H., {Charra}, J., {Grec}, G., {et~al.} 1997, \solphys, 175, 207

\bibitem[{{Goldreich} \& {Keeley}(1977)}]{GK77}
{Goldreich}, P. \& {Keeley}, D.~A. 1977, \apj, 212, 243

\bibitem[{{Goldreich} \& {Kumar}(1990)}]{Gold90}
{Goldreich}, P. \& {Kumar}, P. 1990, \apj, 363, 694

\bibitem[{{Goldreich} {et~al.}(1994){Goldreich}, {Murray}, \& {Kumar}}]{GK94}
{Goldreich}, P., {Murray}, N., \& {Kumar}, P. 1994, \apj, 424, 466

\bibitem[{{Gough}(1977)}]{Gough77}
{Gough}, D.~O. 1977, \apj, 214, 196

\bibitem[{{Gryanik} \& {Hartmann}(2002)}]{GH2002}
{Gryanik}, V.~M. \& {Hartmann}, J. 2002, Journal of Atmospheric Sciences, 59,
  2729

\bibitem[{{Jim{\'e}nez-Reyes} {et~al.}(2003){Jim{\'e}nez-Reyes},
  {Garc{\'{\i}}a}, {Jim{\'e}nez}, \& {Chaplin}}]{Jim03}
{Jim{\'e}nez-Reyes}, S.~J., {Garc{\'{\i}}a}, R.~A., {Jim{\'e}nez}, A., \&
  {Chaplin}, W.~J. 2003, \apj, 595, 446

\bibitem[{{Kraichnan}(1957)}]{Kraichnan57}
{Kraichnan}, R.~H. 1957, Physical Review., 107, 1485

\bibitem[{Lesieur(1997)}]{Lesieur97}
Lesieur, M. 1997, Turbulence in fluids (Kluwer Academic Publishers)

\bibitem[{{Osaki}(1990)}]{Osaki90}
{Osaki}, Y. 1990, in Lecture Notes in Physics : Progress of Seismology of the
  Sun and Stars, ed. Y.~{Osaki} \& H.~{Shibahashi} (Springer-Verlag), 75

\bibitem[{{Rieutord} \& {Zahn}(1995)}]{RZ95}
{Rieutord}, M. \& {Zahn}, J.-P. 1995, \aap, 296, 127

\bibitem[{{Samadi} \& {Goupil}(2001)}]{Samadi00I}
{Samadi}, R. \& {Goupil}, M.~. 2001, \aap, 370, 136

\bibitem[{{Samadi} {et~al.}(2001){Samadi}, {Goupil}, \&
  {Lebreton}}]{Samadi00II}
{Samadi}, R., {Goupil}, M.~., \& {Lebreton}, Y. 2001, \aap, 370, 147

\bibitem[{{Samadi} {et~al.}(2006){Samadi}, {Kupka}, {Goupil}, {Lebreton}, \&
  {van't Veer-Menneret}}]{Samadi05b}
{Samadi}, R., {Kupka}, F., {Goupil}, M.~J., {Lebreton}, Y., \& {van't
  Veer-Menneret}, C. 2006, \aap, 445, 233

\bibitem[{{Samadi} {et~al.}(2003{\natexlab{a}}){Samadi}, {Nordlund}, {Stein},
  {Goupil}, \& {Roxburgh}}]{Samadi02II}
{Samadi}, R., {Nordlund}, {\AA}., {Stein}, R.~F., {Goupil}, M.~J., \&
  {Roxburgh}, I. 2003{\natexlab{a}}, \aap, 404, 1129

\bibitem[{{Samadi} {et~al.}(2003{\natexlab{b}}){Samadi}, {Nordlund}, {Stein},
  {Goupil}, \& {Roxburgh}}]{Samadi02I}
{Samadi}, R., {Nordlund}, {\AA}., {Stein}, R.~F., {Goupil}, M.~J., \&
  {Roxburgh}, I. 2003{\natexlab{b}}, \aap, 403, 303

\bibitem[{{Stein}(1967)}]{Stein67}
{Stein}, R.~F. 1967, Solar Physics, 2, 385

\bibitem[{{Stein} \& {Nordlund}(1998)}]{Stein98}
{Stein}, R.~F. \& {Nordlund}, A. 1998, \apj, 499, 914

\bibitem[{{Stein} \& {Nordlund}(2001)}]{Stein01B}
{Stein}, R.~F. \& {Nordlund}, {\AA}. 2001, \apj, 546, 585

\bibitem[{{V{\"o}gler} {et~al.}(2005){V{\"o}gler}, {Shelyag}, {Sch{\"u}ssler},
  {Cattaneo}, {Emonet}, \& {Linde}}]{Voegler05}
{V{\"o}gler}, A., {Shelyag}, S., {Sch{\"u}ssler}, M., {et~al.} 2005, \aap, 429,
  335

\bibitem[{{Weiss} {et~al.}(2002){Weiss}, {Proctor}, \& {Brownjohn}}]{Weiss02}
{Weiss}, N.~O., {Proctor}, M.~R.~E., \& {Brownjohn}, D.~P. 2002, \mnras, 337,
  293

\end{thebibliography}

\end{document}